# Forecasting e-scooter substitution of direct and access trips by mode and distance

**Mina Lee[1], Joseph Y. J. Chow[*,1], Gyugeun Yoon[1], Brian Yueshuai He[2]**

[1]C2SMART University Transportation Center, New York University Tandon School of Engineering, Brooklyn, NY, USA

[2]Institute of Transportation Studies, University of California Los Angeles, Los Angeles, CA, USA

**ABSTRACT**

An e-scooter trip model is estimated from four U.S. cities: Portland, Austin, Chicago and New York City. A log-log regression model is estimated for e-scooter trips based on user age, population, land area, and the number of scooters. The model predicts 75K daily e-scooter trips in Manhattan for a deployment of 2000 scooters, which translates to 77 million USD in annual revenue. We propose a novel nonlinear, multifactor model to break down the number of daily trips by the alternative modes of transportation that they would likely substitute based on statistical similarity. The model parameters reveal a relationship with direct trips of bike, walk, carpool, automobile and taxi as well as access/egress trips with public transit in Manhattan. Our model estimates that e-scooters could replace 32% of carpool; 13% of bike; and 7.2% of taxi trips. The distance structure of revenue from access/egress trips is found to differ from that of other substituted trips.

# 1. INTRODUCTION

"Micromobility services" is a relatively new term (in the context of urban mobility) defined to encapsulate the set of small vehicle shared mobility modes including electric scooters (e-scooters), docked and dockless shared bikes, electric skateboards, electric mopeds, and electric pedal-assisted (pedelec) bikes (see Zarif et al., 2019).

The bike/e-bike sharing market has grown continuously and the system began to spread worldwide in the early 2010's. Citi Bike, based in New York, started in 2013 and its daily ridership record exceeded 100K trips in a single day in 2019 (Citi Bike, 2019). Another recent growing service is the e-moped market. A New-York based startup, Revel, started its service in 2018. Revel now operates 6K electric e-mopeds in New York, Florida, Washington D.C, and California. (Bloomberg, 2021). Sometimes the term "e-scooter" is used interchangeably with "e-mopeds" services such as Revel. However, in this study we define e-scooters strictly as motorized stand-up scooters.

The e-scooter sharing system was introduced in 2017 as a new shared mobility service in the United States (U.S.) and is now one of the fastest emerging micromobility services. As of 2018, such e-scooter sharing companies as Lime and Bird operate in over 100 cities around the world and have each been valued at over 1 billion USD in addition to receiving hundreds of millions in funding from venture capitalists (Toll, 2018).

The growing popularity of e-scooters can be accredited to its convenient dockless system and small form factor for making short trips. Users only need to download a smartphone application and locate a nearby scooter to start the ride. After reaching their destination, users can park the scooter anywhere, instead of having to return to the original pickup location. This dockless system has made the rental process smoother, enticing prospective users to test out the technology and encouraging conversion to the service (Irfan, 2018).

The e-scooter sharing system has expanded so rapidly not only because of its low entry barrier, but also because e-scooters are potentially filling a mobility gap in cities that have weaker public transit infrastructure (Smith and Schwieterman, 2018). E-scooters can provide better access to transit access points and offer an economical means to travel short distances as part of a Mobility-as-a-Service (MaaS) system (Djavadian and Chow, 2017a). Furthermore, e-scooters offer environmental advantages by reducing traffic congestion and fuel use, which can be a catalyst for adoption in cities where automobiles are the most common mode of transportation.

Considering successful rollouts in major cities like Los Angeles, San Francisco, and Washington D.C., New York City (NYC) is poised to be a major market for e-scooter sharing services as millions of New Yorkers rely on public transit systems every day, with over 2.5 billion trips made in 2018 (MTA, 2018) that can be complemented with e-scooter access. Posed as a solution to the 'last-mile' problem (Djavadian and Chow, 2017b), e-scooter services are expected to not only expand accessibility to transportation services throughout the city but also alleviate traffic congestion (Wondinsky, 2018). Similar micromobility services like Citi Bike (Xu and Chow, 2020) have already created niche markets to serve in Manhattan: the average travel time of Citi Bike rentals within 1.5 miles is 5 minutes faster and less expensive than taking a taxi (NYC, 2016).

While the legality of e-scooters is varied throughout the U.S., growing demand has led many cities to now consider legalization. For instance, in response to thousands of requests from the public, cities like Portland, OR, and Baltimore, MD, designed and implemented e-scooter pilot programs to test the potential impact on reducing pollution and congestion within their urban areas. In Portland, the promising results of the initial pilot program organized by the Portland Bureau of Transportation (PBOT) prompted a second phase trial to improve public safety and develop recommendations for permanent use (PBOT, 2018). Chicago proposed a second pilot program following the first trial in 2019 (BCDOT, 2019). Likewise, the NYC City Council has been challenged by the e-scooter market. As of April 2019, several bills were proposed to legalize e-scooter services in NYC (Fitzsimmons, 2019). As of August 2020, dozens of communities have already started embracing e-scooter systems. Figure 1 shows the cities with e-scooter fleets as orange circles in the interactive map from the Bureau of Transportation Statistics (BTS) compared to 2017 (BTS, 2020).

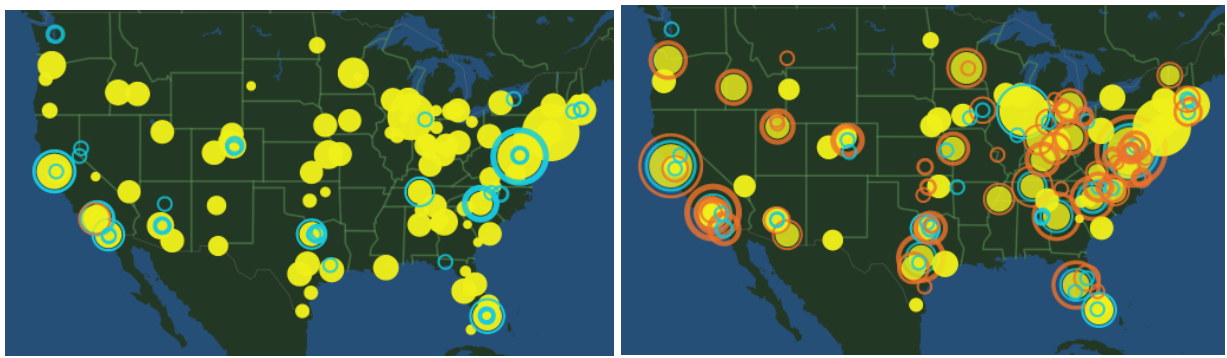

**FIGURE 1.** Interactive bikeshare (dockless: blue, docked: yellow) and e-scooter (orange) map in 2017 and 2020 (BTS, 2020).

With such growth and interest from the private sector, policymakers clearly have a need to forecast the potential demand for e-scooter trips. While predicting the demand for trips from sociodemographic and trip data can be straightforward (e.g. Caspi et al., 2020), using that trip demand to explain substitution from existing trips is more elusive. The problem is that e-scooter trips tend to be shorter in distance, and many may not even be used to substitute an entire trip from origin to destination. Instead, e-scooters may be used to replace an access/egress trip to a public transit stop.

This study addresses this research challenge and gap in the literature by *forecasting the potential modal substitution* of e-scooter deployment in Manhattan. Three contributions are made. First, the e-scooter trip demand in Manhattan is forecasted using a new trip generation model estimated from data from e-scooter pilot programs in three different cities. By using data from three cities, we can include fleet size as an explanatory variable. Second, we propose a novel, nonlinear, distance-based factor model that distributes that demand into different fixed fractions of each mode by distance based purely on statistical similarity. The model separates e-scooter trips into two types: competitive substitutions (direct trips) and complementary substitutions (access trips). Lastly, the model is fit to the existing modal trip landscape in Manhattan (using data from a synthetic population constructed in a prior study, He et al., 2020) to uncover the modes most similar to e-scooters and to determine distributions of e-scooter trips by distance, by access or direct trip substitution, and to forecast corresponding revenues.

## 2. LITERATURE REVIEW

Although e-scooters have become increasingly popular in metropolitan cities, there exists limited research due to the relative infancy of e-scooter data, in contrast to ubiquitous bike sharing data. Most research with e-scooter ridership data analyzed patterns from observed e-scooter ridership data and the characteristics of riders. Degele (2018) pointed out that e-scooter riders in Germany can be clustered by four customer segments based on their demographic characteristics and trip attributes. Smith and Schwieterman (2018) analyzed the patterns of e-scooter trips in Chicago and demonstrated the potential impact of e-scooters. Noland (2019) studied e-scooter data from Louisville, Kentucky. Other studies have been conducted in multiple cities by their departments of transportation through pilot programs (PBOT, 2018; BCDOT, 2019). A more environmental-based analysis is presented about the positive effects of e-scooters on congestion and pollution in cities (Hardt and Bogenberger, 2019). Zou et al. (2020) examined the travel patterns observed on e-scooters in Washington, DC. None of these studies above have estimated forecast models for e-scooter ridership.

To estimate demand in different areas of transportation, trip generation regression models (McNally, 2007) are useful tools because of their simplicity. To quantify relationships between demographics and bike sharing ridership, some studies aggregate demographic variables for each spatial area (Ranaiefar and Rixey, 2015). Munira (2017) showed that micromobility demand tends to be attributed to several demographic and socioeconomic factors. Capital Bikeshare (2012) discovered that demographic characteristics such as employment rate, education levels, average age, and gender can be categorized in their user memberships. Hankey and Lindsey (2016) demonstrated that pedestrian and bicycle traffic are mostly attributed to district characteristics like high accessibility to jobs. Through this demographic perspective, clustering analysis regarding socio-demographic properties and travel modes was suggested to improve public bike systems (Geng et al., 2016). Likewise, a study about e-scooter sharing systems also considered the usage patterns of e-scooters and the behavioral characteristics of riders. Observed e-scooter ridership data from pilot programs in Portland and Baltimore are highly correlated to groups between 20-40 years of age (PBOT, 2018; BCDOT, 2019). McKenzie (2019) and Hosseinzadeh et al. (2020) related e-scooter trips in one city to different built environment attributes. Bai and Jiao (2020) used two cities' data for a similar purpose. Caspi et al. (2020) developed a spatial trip prediction model for e-scooters but did not relate those trips to substitution from other existing modes. Reck et al. (2021) focused on competition between different micro-mobility options, but not other modes.

Spatial correlation of travel behavior has been studied for demand predictions (e.g. Sener et al., 2011; Lin et al., 2018; Yu and Peng, 2019). In this study, however, we exclude this factor since it prioritizes studying the substitution of existing travel modes by e-scooter, not measuring the degree of spatial correlation of them. Hosseinzadeh et al. (2021) already compared a global ordinary least squares model with a geographical weighted regression when analyzing the relationship between the density of e-scooter trips and zone-level factors such as land use and age distribution.

City mobility options include carpool, taxi, bike, walking, driving and public transit. If a new travel mode becomes available, mode substitutions may occur. Recent research has looked at the impact of a new mode like ride hailing on other modes (Gerte, 2019; Jin, 2018). Some studies in travel behavior analysis relate socio-economic factors which can influence the transition from existing modes to include the new micromobility service. For example, Yang et al. (2016) found that current bike sharing users were often converted from existing nonmotorized transportation modes when a bike sharing program is introduced. Another study also claims that the choice of travel between bike, automobile, and public transit is determined by sociodemographic characteristics and one's context (Yang et al., 2015). Likewise, Smith and Schwieterman (2018) discovered that e-scooters would be a strong alternative to private automobile trips for short distances (between 0.5 and 2 miles). This supports Portland's pilot findings that 34% of Portlanders would have chosen motorized travel modes if e-scooters were not available (PBOT, 2018).

Due to convenience and ease of accessibility, micromobility services are often combined with other travel modes in the main trip. With the success of bike sharing, bike-and-ride can be used to substitute other access modes from and to public transit stations (Doolittle and Porter, 1994; Shaheen, 2010). Travel choice models estimated by Fan et al. (2019) for first/last mile trips with bike sharing systems found that bike sharing can take 45.9% of mode share for first/last mile trips. They considered differences in choice behavior between various age groups and found young adults differed from middle-aged individuals. Halldórsdóttir (2017) found that mode choices of access/egress in Copenhagen are related to the travelers' characteristics and the purpose of the trip.

Thus far, e-scooter demand modeling has been limited to prediction of direct trips via techniques like cosine similarity and Moran I statistic for spatial analysis (McKenzie, 2019); and for trip count prediction, a negative binomial regression model (Bai and Jiao, 2020) and geographically weighted regression model (Caspi et al., 2020). There has not been a study that relates e-scooter trips to different existing modes like taxi and public transit, nor by distance of trip or by direct versus access/egress, and none that considers data from three or more cities that capture variation in fleet size.

## 3. DATA AND DATA PROCESSING

### 3.1. E-scooter ridership data

We start with data from the e-scooter pilot programs that are arranged in three different cities: Portland, Austin, and Chicago. PBOT organized an initial four-month pilot program from July to November in 2018. The program has published an analysis report along with the survey data collected from e-scooter riders who had participated during this time. The survey results inform the socio-demographic data of the participants, including income, age, purpose of the ride, and alternative mode substituted by the trip (PBOT, 2018). The Portland survey data provides a background regarding the riders and the characteristics of the trips: 30% of trips were for commuting, whereas 28% were for pleasure or recreation. Some riders used e-scooters to substitute for other modes of transportation, such as driving or walking. Riders in the age group of 20-50 made up 86% of Portland e-scooter riders. For the duration of the pilot period, the official total ridership of e-scooters in Portland was 700K (PBOT, 2018). The publicly available e-scooter ridership data mined from Portland are street count data obtained from routes traveled by e-scooter riders. To determine the ridership by zip code of origin, the 700K total trips were distributed across the 28 zip codes by mapping out the percentage of riders traveling within each zip code.

Austin, Texas, started its own pilot program that ran between August and November 2018. The Austin Transportation Department published a community survey report, which stated that Austin residents, like their Portland counterparts, frequently used micro-mobility services for commuting, recreation, and errands (CATD, 2019). Users also consider the micro-mobility service as a substitute of walking, ride-hailing, carpooling and bicycle. Lastly, Chicago organized the e-scooter pilot program between June and October in 2019, and 821K e-scooter trips were reported. Chicago users rode e-scooters most frequently during the evening rush period on weekdays and during the afternoon period on weekends (CDOT, 2020). For the ridership demand in Chicago and Austin, we use public ridership data collected at a community level. After converting the community level ridership into individual zip-codes by mapping out the percentage of intersecting areas, the ridership numbers in Chicago and Austin are recorded in 14 and 36 zip codes, respectively. This total ridership for each city is broken down into daily trips depending on the duration of the pilot program, which provides the average daily ridership.

By collecting data from various cities and time periods, we create a general model that considers the attributes of different cities. Even though each city provided their e-scooter ridership data in different formats, we convert them into zip code level data based on the origin of those trips. This results in 78 observations. For the modeling, we remove the records that have fewer than 10 daily e-scooter rides in a zip code to remove the outliers, leading to 60 records in total out of the 78 original ones. A model estimated from this data is introduced in Section 4.

### 3.2. Sociodemographic and Traffic Data

We use the 2017 American Community Survey (ACS, 2018) to obtain demographic and economic statistics based on zip code (Towncharts, 2018). A variety of factors are collected as potential independent variables. Some of the data is summarized in Table 1. Portland, Austin, and Chicago are used for estimating an e-scooter model while NYC is used for applying the model. Among the 78 zip codes, only 60 are kept after removing outliers as mentioned in section 3.1.

**TABLE 1.** Demographic statistics of each zip code in Portland, Texas, Chicago, and New York

| Cities | | Population density (/sqmi) | Population | Income ($) | Land Area (sqmi) | Age Ratio (%) | Unemployment Rate (%) | Laborforce Participation (%) | Health Insurance (%) |
|---|---|---|---|---|---|---|---|---|---|
| For estimating e-scooter forecast model | | | | | | | | | |
| Portland (N=28) | Avg | 5,803 | 28,152 | 61,998 | 9.71 | 49.5 | 4.6 | 69 | 89.7 |
| | Min | 62 | 4,396 | 32,276 | 0.25 | 29.8 | 2 | 59 | 84.1 |
| | Max | 15,785 | 50,320 | 94,192 | 64.33 | 63.6 | 9 | 79 | 96.5 |
| Austin (N=36) | Avg | 3,685 | 31,212 | 75,316 | 18.51 | 49.6 | 2.9 | 71.7 | 86.6 |
| | Min | 98 | 1,815 | 19,188 | 1.51 | 26.4 | 1 | 45 | 71.3 |
| | Max | 15,462 | 89,830 | 150,000 | 104.97 | 73.7 | 4 | 82 | 98.3 |
| Chicago (N=14) | Avg | 15,792 | 62,544 | 64,308 | 4.12 | 51 | 4.9 | 68.3 | 84.4 |
| | Min | 10,229 | 20,201 | 22,992 | 1.7 | 38.9 | 2 | 50 | 74.9 |
| | Max | 22,521 | 94,395 | 107,461 | 7.1 | 70.8 | 9 | 85 | 95.8 |
| For application of e-scooter forecast model | | | | | | | | | |
| New York (N=28) | Avg | 81,009 | 37,317 | 100,189 | 0.5 | 52.4 | 3.9 | 70.1 | 93.1 |
| | Min | 5,484 | 1,324 | 27,932 | 0.01 | 28.6 | 1 | 51 | 84.3 |
| | Max | 159,898 | 94,382 | 250,001 | 1.45 | 80.5 | 10 | 90 | 99 |

Since the basic geographic zonal system of NYC data from The New York Metropolitan Transportation Council (NYMTC) and The New York Best Practice Model (NYBPM) is Transportation Analysis Zones (TAZ), we choose to use TAZs for this research analysis to be consistent with the major data sources. We restrict the study area to districts in Manhattan which fall under TAZ 1 to 318 (N = 318). NYMTC 2010/2011 Regional Household Travel Survey (RHTS) (NYMTC, 2011) collected sample trip data from the public in the New York Metropolitan Area. While the modes have since expanded further, a validation study from He et al. (2020) comparing with the 2017 NYC Citywide Mobility Survey (CMS) found that they did not change that significantly population-wide.

This data provides origin-destination demand based on TAZ level and six different trip modes. The origin-destination based trip data is converted both to trip distances and travel times for the model designed in Section 4.2. Trip distances for all modal trips except public transit are determined from the 2018 LION (NYC, 2018) road network. The public transit trip travel times are determined using a transit router function (Balac, 2018) in MATSim with General Transit Feed Specification (GTFS) data from 2016 (TransitFeeds, 2019). The transit travel time includes egress, access, and transfer time.

There were 5.842 million observed total daily trips in Manhattan in 2011, and the mode shares of carpool, transit, taxi, bike, walk, and auto were 2%, 29%, 4%, 2%, 60%, and 4%, respectively. As we can see, Manhattanites traveling in the same borough mostly walk, with only a very low percentage driving a car.

Lastly, for the Citi Bike data, we obtain total trip counts from their website which publishes daily ridership data, station locations, and travel times between stations (Citi Bike, 2018). The locations of origin/destination stations are converted to TAZ level using QGIS. Some districts do not have access to Citi Bike since there are no stations nearby. For this research, the average daily ridership within Manhattan is collected based on the locations of origin stations with the same period of Portland and Austin's pilot program (July to November 2018). The total number of daily Citi Bike trips is 45K. The modal trip counts in Manhattan are summarized in Table 3 and by distance in Figure 3.

**TABLE 3.** Modal trip counts based on the zone of origin in Manhattan

| Mode | Data source | Average Daily Count by TAZ | Count Standard Deviation by TAZ |
|---|---|---|---|
| Carpool | 2011 RHTS | 296 | 805 |
| Public transit | 2011 RHTS | 5,314 | 4,937 |
| Taxi | 2011 RHTS | 825 | 1,556 |
| Bike | 2011 RHTS | 350 | 693 |
| Walk | 2011 RHTS | 11,162 | 10,730 |
| Auto | 2011 RHTS | 661 | 1,446 |
| Citi Bike | Citi Bike (2018) | 122 | 157 |

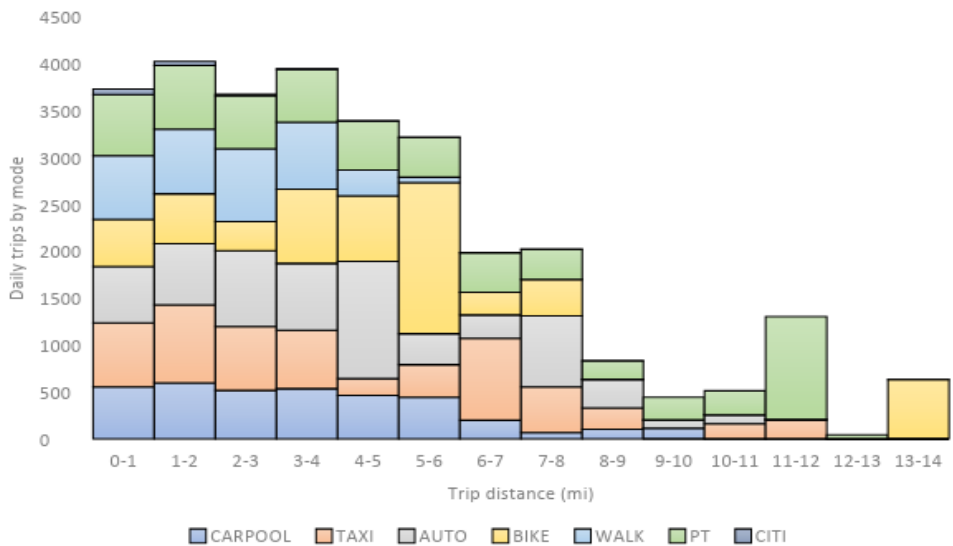

**FIGURE 3**. The number of daily trips per mode by travel distance.

## 4. METHODOLOGY

Because e-scooter ridership data in New York are not readily available to the public, the research design involves a two-step approach: (1) estimate and apply a demand model to forecast the potential demand in different zones within Manhattan; (2) estimate a nonlinear multifactor model to break down that demand into different modal trips such that statistically-significant factors would emerge. For example, based on the distribution of Citi Bike and taxi trip data, is it likely that e-scooters could be statistically explained by 2% of the former and 4% of the latter? How much would these factors be affected by travel distance? The model is not meant to make behavioral predictions of which modes would be substituted by e-scooter. The framework for the methodology is highlighted in Figure 2.

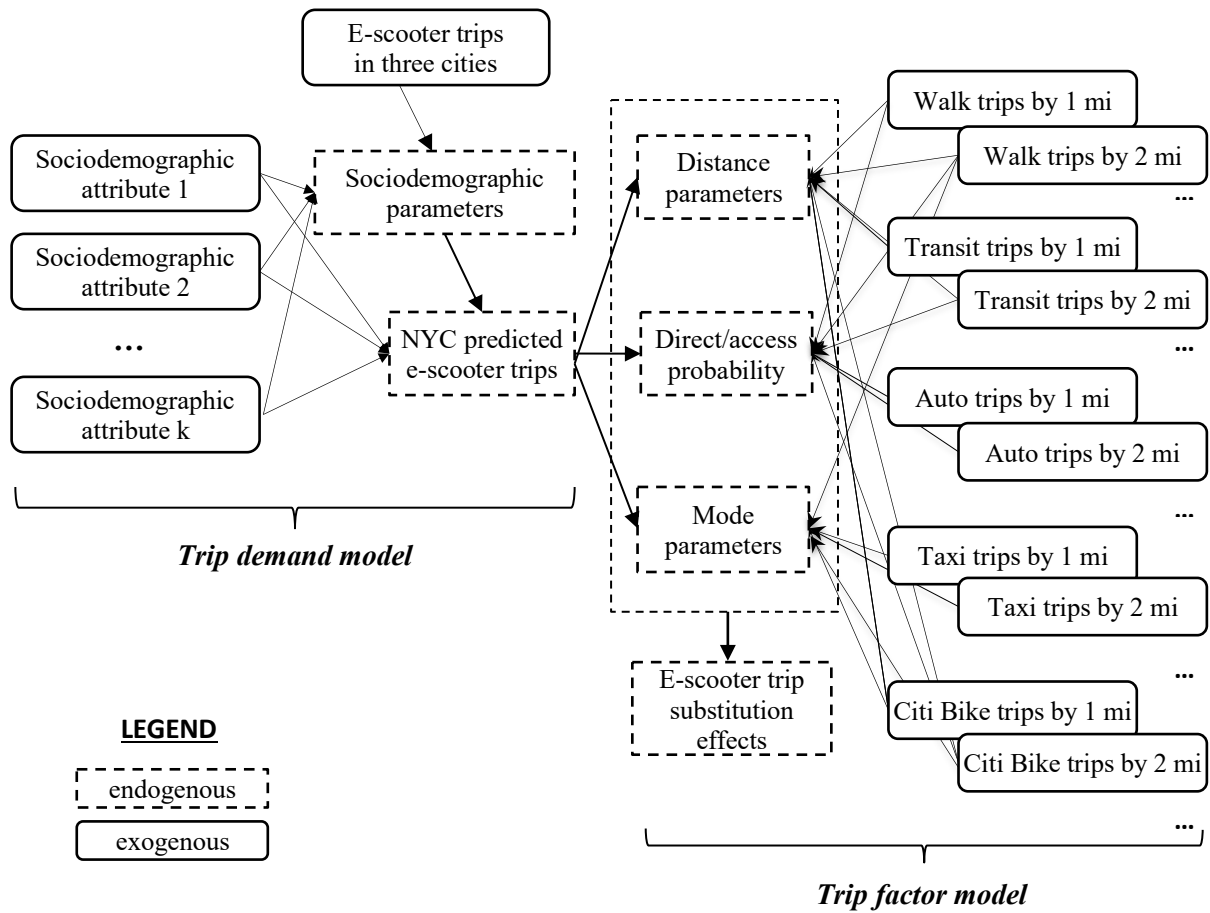

**FIGURE 2.** Data Flow Framework for research design.

The demand regression model is estimated from demographic attributes such as population and employment because they tend to have a positive relationship with nonmotorized activity (Munira, 2017). A nonlinear regression model is then estimated with modal trips as independent variables and the forecasted e-scooter demand as the dependent variable. The model multiplies the modal trips with a probability that varies by distance. The probability represents the share that e-scooter ridership could substitute from existing modes; as specified in the model, it could either draw from direct trips or from the access modes to public transit. Estimation of the model then reveals the modal trips that could contribute to e-scooter demand, the amount that they draw from, and how that amount varies by distance of the trip.

**4.1 Multivariate log-linear regression model for forecasting e-scooter trip demand estimation**

To estimate the e-scooter trip forecast model, the social, economic, and demographic data collected from the survey are used. Several demographic variables are considered, as shown in Table 4. With sociodemographic variables, the demand of e-scooter ridership ($R_{est,i}$) is predicted by zip code of trip origin.

**TABLE 4.** Potential independent variables in a coverage area

| Variables | Description (units) |
|---|---|
| Income | The median household income ($) |
| Labor | The average rate of labor force participation (%) |
| Population | The number of people living in a zipcode |
| Unemployment | The average rate of unemployment (%) |
| Population Density | The population divided by a land area (/sqmi) |
| Age Ratio | The ratio of ages 20 to 49 over the total population (%) |
| Health Insurance | The average percentage of health insurance holders (%) |
| Land Area | The land area (sqmi) |
| Scooters | The number of scooters provided in the whole service area* |

* The scale unit is the total service area. Other variables are collected at a zip code level.

We test nonlinear models, such as exponential and polynomial models, as well as linear ones. The 60 zip codes from the three cities were split 80% (48 records) for estimating the model and the remaining 20% (12 records) are randomly selected from the three cities to be withheld as an out-of-sample test set. After applying different combinations of independent variables in conjunction with different models, we finally choose a multiple log-log regression model with four variables. This log-log linear demand function is well-known to be suitable for econometric purposes, since it captures the elasticities of each variable from the parameters of a driven demand model (Domencich et al., 1968; Dritsakis, 2000). The parameters are estimated using ordinary least squares conducted in $R$, the statistical computing software.

$$ln(R_{est,i}) = \beta_{const} + \beta_P \cdot ln(Population * Age\ Ratio) + \beta_L \cdot ln(Land\ Area) + \beta_S \cdot ln(Scooters) + \varepsilon_i \quad (1)$$

where

$R_{est,i}$ is the total number of e-scooter trips generated in zip code $i$

$\beta_k$ is the constant and coefficients of the attributes, $k = \{const, P, L, S\}$

The estimation results are shown in Table 5. All the coefficients are found to be statistically significant and the $R^2$ of the model is 0.314, which is an adequate fit considering the nonlinearity. A q-q plot and residual plot shown in Figure 4 further indicate the good fit of the model.

**TABLE 5.** Summary of Statistics for log-log multivariate regression model

| Variables | Coefficient | Std. Error | t.value | p-value |
|---|---|---|---|---|
| $\beta_{const}$ | -7.929 | 4.6446 | -1.707 | 0.0949* |
| $\beta_P$ | 0.573 | 0.2519 | 2.276 | 0.0278** |
| $\beta_L$ | -1.108 | 0.2690 | -4.117 | 0.0002*** |
| $\beta_S$ | 0.7812 | 0.2966 | 2.634 | 0.0116** |

*, **, and *** indicate statistical significance at the 0.1, 0.05, 0.01 levels, respectively.

| Summary of Fit | |
|---|---|
| Residual standard error: 1.416 | p-value: 0.0008007 |
| Multiple R-square: 0.3136 | Observations: 48 (Degree of Freedom: 44) |
| Adjusted R-squared: 0.2668 | F-statistic: 6.7 |

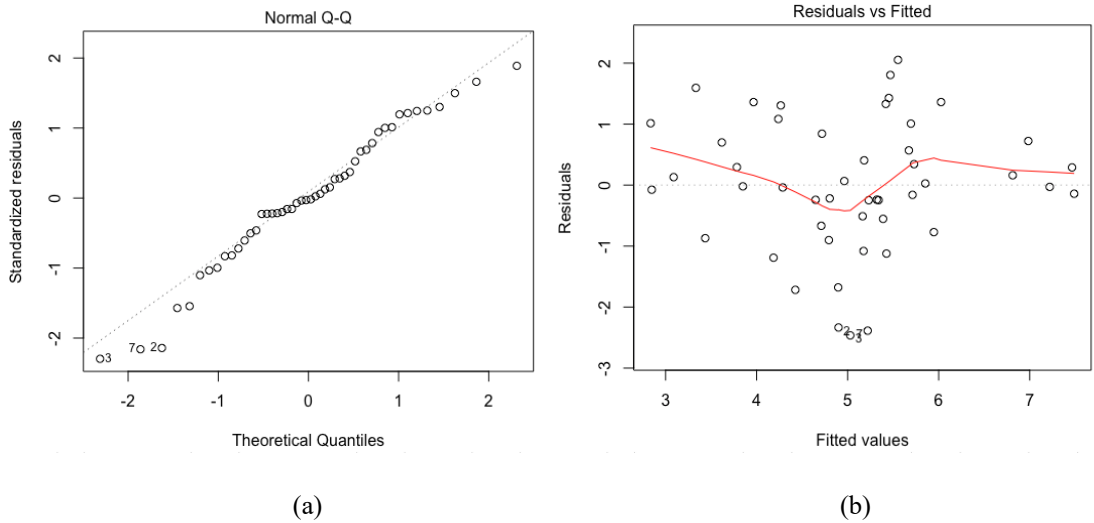

(a)            (b)

**FIGURE 4.** (a) q-q plots of the e-scooter ridership by zip code in Portland and (b) Residual plots.

Age group is one of the key factors of e-scooter demand, and this analysis duplicates the results of the survey in Portland, showing that the targeted riders (ages 20 through 50) constitute a significant portion of total ridership. Since e-scooter service is only operated by smartphone app, the requisite ownership of a smartphone can affect the usage of e-scooters. The percentage of smartphone ownership in the targeted age group has increased to around 90% from 2016 to 2018, whereas the age group of 50+ remains similar at 74% or less (Pew Research Center, 2018).

Additionally, we consider the accessibility and availability (the supply side) of the e-scooters as an explanatory variable to the forecast model, only possible because of the use of multiple cities' data. Each pilot program provided a different number of e-scooters, and the sizes of the areas in which the programs were conducted varied. Therefore, the number of scooters and the square mileage of the program's coverage area could affect the accessibility of e-scooters. For this analysis, we choose to use a fleet size of 2000 as an input for the model in NYC. This fleet size is chosen by comparing to other pilot programs based on the service area and the population. The number of dispatched scooters in the other cities with pilot programs were 2K in Portland; 7K in Austin; and 2.5K in Chicago. Each city's fleet size per service area (sqmi) was about 44 in Chicago; 8.7 in Portland; and 5.8 in Texas. In New York, the 2000 fleet supply provides 88 e-scooters per land area, and respectively 131 e-scooters based on a fleet supply of 3000. We keep a similar level to the 44 in Chicago considering the similar city infrastructure and transportation system. If we increase the total number of e-scooters from 2000 to 3000, the e-scooter demand is projected to rise since the fleet size is positively related to the ridership.

All variables are substantially significant to explain the model with p-values less than 0.05. In a log-log regression model, the coefficient of each variable represents an elasticity with respect to the dependent variable (see Benoit, 2011). We can use a coefficient to describe the impact of a percentage change in each variable. For instance, the elasticity of the population in the age group of those between 20 and 50 is 0.5, which demonstrates that increasing the population of the 20-50 age group by 10% would increase the e-scooter ridership by 5%.

**4.2 E-scooter regression model validation**

The estimated log regression model in section 4.1 is validated in two ways: (a) with a performance evaluation from a 12-sample out-of-sample test data set and (b) a comparison of the model application to one zip code in Hoboken based on data shared by Lime.

For the model fit, the Mean Absolute Error (MAE) and Coefficient of Variation (CV) are used. Ospina (Ospina et al, 2019) found that CV provides a suitable performance when compared to the classic estimator. Figure 5 shows that the log model performs well with the CV value of 0.27 in predicting the ln(trips). Among 12 test records, 67% of them are within an error range of 20%. In terms of each city's records, Texas data points have relatively higher error bound than other cities. Leaving out the Texas test data, the average disturbance of the ln(trips) falls to within 6%. Since our focus is on NYC, which shares similar urban-environmental features with Chicago, we find this acceptable.

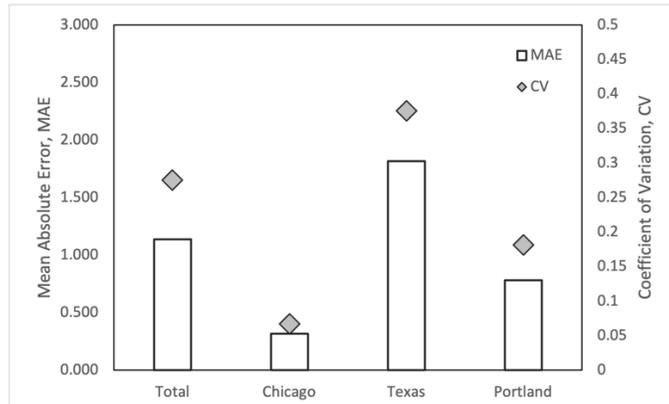

**FIGURE 5.** Mean Absolute Deviation (MAD) and Coefficient of Variation (CV) of test data prediction.

Furthermore, we apply the model to the recent e-scooter pilot program in Hoboken, NJ (just across the Hudson River from Manhattan), which supplied 300 e-scooters (Hoboken Girl, 2019) and had 3.1K average daily trips between May and June. With the estimated value of Hoboken demographic data, the error is 0.42 of the observed value. Given that this is a single observation point (one zip code), the model's performance is satisfactory in predicting the ln(trips).

### 4.3 Conversion of Zip Codes to TAZ

The ridership data estimated through the regression model is based on zip code. Thus, the zip code level data are mapped onto the TAZ level to leverage the NYMTC's household travel survey data for the upcoming nonlinear multifactor model in section 4.4. Some zip code areas overlap with multiple TAZs. We first calculate the proportion of population in each zip code with TAZ level populations. Then, the ridership of each zip code is distributed to TAZ level ridership using the prior proportions.

For instance, suppose zip code 10005 overlaps with TAZ 4, 5, and 6. We would sum up populations of those three TAZ areas and calculate the ratios of each population, which are 13%, 58%, and 29%, respectively. Then, the predicted ridership of zip code area 10005, which is 5820, is distributed to the three TAZ areas based on the prior proportions: 757, 3376, and 1687 trips. Since TAZ areas also overlap with multiple zip code districts, therefore, e-scooter trips from different zip code are added up for the total ridership of each TAZ area.

We use population data instead of the areas of overlapped districts. Since e-scooters are spread irregularly and a TAZ level is small enough for users to walk to neighborhoods, the potential number of users has more relevance than the size of the area.

## 4.4 Nonlinear Multifactor Model

In section 4.1, we find the total daily forecasted e-scooter ridership by TAZ level ($Z = \{1,2,,...,318\}$). However, we do not know which modes the total e-scooter trips would be drawn from, whether it is replacing another direct mode like $M = \{bike, walk, citi\ bike, carpool, taxi, auto, or\ public\ transit\}$, replacing an access/egress mode for public transit, or drawn from latent demand. Therefore, we propose a novel nonlinear regression model as a type of multifactor model that breaks down the e-scooter trip demand into the most statistically significant components. These components are drawn as replacement trips competing against either direct trips for all modes ($N_{m,i}, m \in M, i \in Z$) or access trips using public transit ($m =$

{*public transit*}). The dependent variables are the forecasted e-scooter trips per zone while the independent variables are the existing trips by mode.

In designing the model, we hypothesize that distance should play an important role. Shorter distance trips should be more likely to be replaced by e-scooters. One might intuit that certain modes should have impedances that decrease with distance in e-scooter substitution, such as walk and bike, as e-scooter would be more convenient and faster in speed. However, since e-scooter also includes a variable fare price component, the net effect should likely still be an increasing impedance by distance. For example, some pilot studies already demonstrate the average distance of e-scooter ridership is typically less than 2 miles (PBOT, 2018; BCDOT 2019). Existing e-scooter fee structure scales in proportion to duration, thus the utility of e-scooter choice for longer trips is expected to be less than shorter trips (Smith and Schwieterman, 2018).

To capture this dependency, we propose a parameter $P_d$ that varies by distance category $d \in D$, where $D = \{0-1\ mi, 1-2\ mi, 2-3\ mi, \ldots, 13-14\ mi\}$, which is used to quantify the percent of modal direct trips that are in competition with e-scooters. The modal trips can be further segmented by distance category, $N_{m,i,d}$. We should see this parameter increase when the distances of trips are shorter. We specify the following relationship in Eq. (2), where $\beta_d$ is a calibrated parameter set per distance category, and $\delta_d$ is the average distance of trips made for category $d$. An alternative design is to split the impedance by mode. This design could be more realistic but is more statistically costly.

$$P_d = \frac{\beta_d}{\delta_d} \qquad (2)$$

Given the portion $P_d$ of $N_{m,i}$ that are subject to competition with e-scooters, another parameter set is introduced to represent the proportion $F_m$ of modal trips that would be replaced by e-scooters. This proportion varies by modal trips but is fixed by distance. If the different modes contribute to e-scooter trips differently, then $F_m$ would vary across modes significantly. Lastly, a constant $C$ is used to capture everything that cannot be controlled for: statistically insignificant trip fractions for modes, induced demand, or tourists' trips not captured by the 2011 RHTS trips.

For the trips that are not subject to competition with e-scooters (i.e. $(1 - P_d)N_{m,i}$), they generally do not contribute to e-scooter trips, except for public transit. For the latter trips, e-scooter trips that do not compete directly may still end up substituting the access trips. We propose that the fraction $F'_{pt,i}$ of public transit trips subject to e-scooter competition with the access trips have a functional form shown in Eq. (3) to be estimated, where $t_i^A$ is the average access time (in hours) for trips to public transit from zone $i \in Z$ and $t_i^E$ are the corresponding average egress trip times into zone $i \in Z$.

$$F'_{pt,i} = \beta_0 + \beta_1 * t_i^A + \beta_2 * t_i^E \tag{3}$$

The final nonlinear regression model for statistically attributing the e-scooter ridership to different modes by distance is shown in Eq. (4).

$$R_{sub,i} = C + \sum_{m \in M} F_m \sum_{d \in D} P_d N_{m,i,d} + \sum_{d \in D}(1 - P_d)F'_{pt,i}N_{pt,i,d} + \gamma_i \tag{4}$$

where $\gamma_i$ is the disturbance in the nonlinear regression model, $R_{sub,i}$ are the e-scooter trips per zone $i$ predicted from the model in Eq. (1). Since both $F_m$ and $P_d$ need to be estimated, this model is nonlinear in its parameters. The least-squares regression methodology is applied to estimate the parameters using Excel Solver with the GRG Nonlinear method. Parameters are sought to minimize the gap between the demand for e-scooters and the substituted trips for each zone. The objective function is shown in Eq. (5). The estimated coefficients using the 318 observations in Manhattan assuming a deployed fleet of 2000 scooters are reported in Table 6. Interpretation of these results are presented in Section 5.

$$\min Z = \sum_{i \in Z} (R_{est,i} - R_{sub,i})^2 \tag{5}$$

We use the bootstrap method to evaluate the statistical significance of the estimated coefficients (Chernick, 2014). We sample (with replacement) from the 318 observations to reconstruct 150 bootstrap samples to obtain the confidence intervals for the estimated parameters as well as the standard errors and t-statistics.

## 5. RESULT AND DISCUSSION

### 5.1 Estimation results

The results from the models and analysis are quite insightful. First, having constructed the e-scooter forecast model in section 4.1, we apply it to Manhattan to obtain the number of predicted e-scooter trips per zip code assuming a deployed fleet of 2000 scooters. The model predicts that the total daily e-scooter ridership in NYC under this assumed fleet size would be 75K, which is equivalent to about 1% of total trips within Manhattan.

A large portion of rides are projected to occur in the upper east side and lower Manhattan as shown in Figure 6(a). Citi Bike, which is the most similar mode to e-scooters, had 45K total ridership over the same period in 2018. The estimated e-scooter ridership is approximately 60% higher than that of Citi Bike. The higher number can be explained by the fact that accessibility for Citi Bike is more limited in some areas and dependent on availability of stations and bike lanes (Xu and Chow, 2020).

The difference is clearer when we observe Figure 6(b) showing ridership by e-scooters and Citi Bike by zip code. Generally, e-scooter ridership is spread among the whole Manhattan area. However, in areas where Citi Bike is not accessible (as shown by near zero ridership), e-scooters are not restricted due to their dockless system.

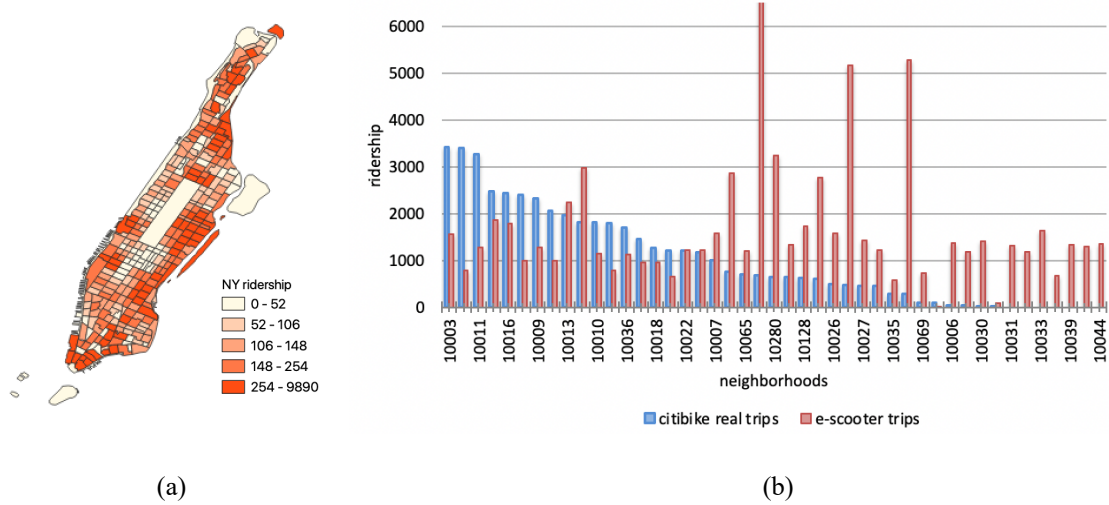

**FIGURE 6.** (a) E-scooter ridership per TAZ and (b) comparison of Citi Bike and e-scooter trips by zip code.

The nonlinear multifactor model breaks the 75K daily e-scooter trips down into two categories: direct and access/egress trips. By structuring the model into direct or access trips while incorporating distance, the results provide an indication of where the trips likely drawn from, in terms of travel modes. The estimation results are shown in Table 6 along with bootstrapping from 150 observations.

**TABLE 6.** Results of parameters for nonlinear regression model

| | | 90% Confidence Interval | | |
|---|---|---|---|---|
| Variables | Coefficient | Lower | Upper | Standard Deviation (t-stat) |
| $Constant(C)$ | 73.365 | 69.999 | 100.001 | 14.013 (5.235)* |
| $F_{bike}$ | 0.226 | 0.000 | 0.307 | 0.093 (2.430)* |
| $F_{walk}$ | 0.021 | 0.000 | 0.201 | 0.060 (0.350) |
| $F_{citi\ bike}$ | 0.094 | 0.000 | 0.208 | 0.086 (1.093) |
| $F_{carpool}$ | 0.636 | 0.000 | 0.700 | 0.264 (2.409)* |
| $F_{taxi}$ | 0.184 | 0.000 | 0.227 | 0.078 (2.359)* |
| $F_{auto}$ | 0.109 | 0.000 | 0.206 | 0.065 (1.677) |
| $F_{public\ transit}$ | 0.000 | 0.000 | 0.108 | N/A |
| $\beta_d$ | 0.493 | 0.400 | 0.500 | 0.037 (13.324)* |
| $\beta_0$ | 0.000 | 0.000 | 0.100 | N/A |
| $\beta_1$ | 0.069 | 0.000 | 0.233 | 0.081 (0.852) |
| $\beta_2$ | 0.000 | 0.000 | 0.050 | N/A |
| Number of observations (TAZs): | | 318 | | |
| Number of bootstrap samples: | | 150 | | |

* statistically significant at 5% level

## 5.2 Analysis of estimated parameters

Certain proportions of each mode can be substituted by e-scooters with some statistical certainty, as shown in Table 6. For instance, based on the $F_{taxi} = 0.184$, we can conclude that implementing a Manhattan-wide e-scooter program with 2000 scooters results in trips that can be statistically explained by substitution of up to 7% ($P_d \times F_{taxi}$) of taxi trips. The carpool trips ($F_{carpool} = 0.636$) are about three times more than the bike trips ($F_{bike} = 0.226$) and taxi trips ($F_{taxi} = 0.184$). The trips with direct public transit by themselves are not statistically significant regarding the e-scooter replacement, likely because the public transit trips exhibit nonlinear distance impedances (the number of very short public transit trips are low compared to longer distance trips – they tend to peak at a further distance than e-scooter trips would be used for). Based on the estimation, the access/egress factor becomes Eq. (6). As a result, up to 24% of the access/egress trips to a public transit station could be substituted with e-scooter trips.

$$F'_{pt,i} = \beta_1 t_i^A \tag{6}$$

The statistically significant value $\beta_d = 0.493$ further suggests that e-scooters are likely to compete with other modes at shorter distances than at longer distances. This is further confirmed by the positive values of all the mode factors except for direct public transit trips, which do not fit into this structure and likely requires a nonlinear distance structure. As shown in Table 7, the competition metric $P_d$ ($\beta_d/\delta_d$) drops by an order of magnitude when the distance increases from 0.5 ($P_d = 0.986$) to 5.5 ($P_d = 0.09$).

**TABLE 7.** Results of the parameter ($P_d$) by average distance

| Distance($\delta_d$) | 0.5 | 1.5 | 2.5 | 3.5 | 4.5 | 5.5 | 6.5 | 7.5 | 8.5 | 9.5 | 10.5 | 11.5 |
|---|---|---|---|---|---|---|---|---|---|---|---|---|
| $P_d$ | 0.986 | 0.329 | 0.20 | 0.14 | 0.11 | 0.09 | 0.08 | 0.07 | 0.06 | 0.05 | 0.05 | 0.04 |

The constant $C = 73.365$ aggregates unexplainable impacts such as latent demand and tourist trips. The constant itself explains 23K of the e-scooter trips. Among the seven different mode choices, carpool, bike, and taxi trips tend to be significantly substituted. Assuming the comparable travel speeds to bikes (10 mph), e-scooters can be an effective replacement for cars, just like e-bikes (Kroesen, 2017). With the estimation, e-scooters can replace, in total, up to 32% of carpool, 13% of bike, 7.2% of taxi, 1.9% of walking, and 1.8% of auto trips, but a majority of these replacements come from short distance trips. The distribution of substituted modes by distance is shown in Figure 7. The shorter the trip distance is, the more likely a trip will be replaced by e-scooters.

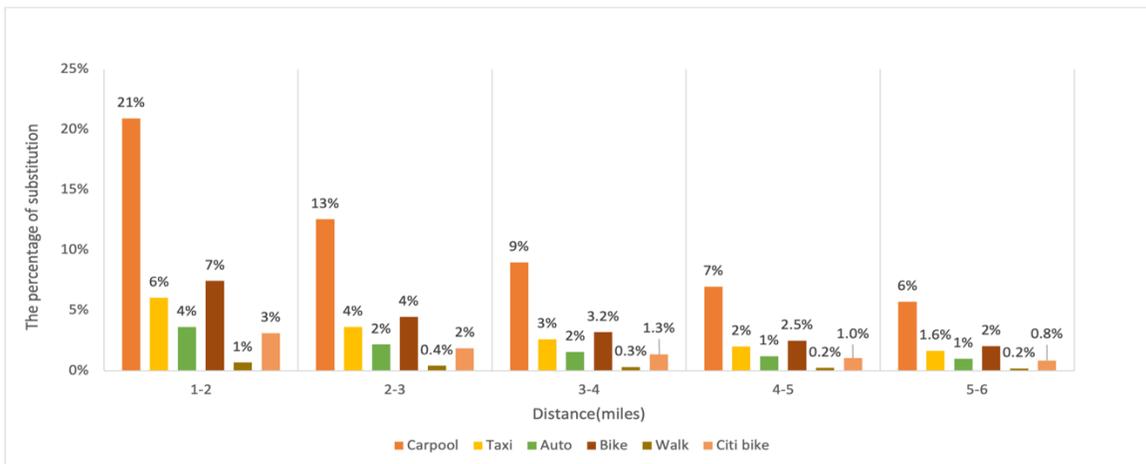

**FIGURE 7.** The percentage of substitution of each mode's trips by distance (in miles).

## 5.3 Spatial modal analysis

We also analyze the distributions of substituted trips for each TAZ by summing up all distances traveled (Figure 8). We see clusters concentrated around Lower Manhattan and the Upper Westside (west of Central Park) for the bike substitution. Most of the carpool replacements take place in Northern Manhattan. For the substituted taxi trips, it is relatively scattered around Manhattan, unlike the other modes.

These spatial distributions provide suggestions of how deployments in various neighborhood can have differing impacts on existing modes. For example, e-scooters deployed primarily in Tribeca/City Hall (bottom segment of Manhattan) would be drawing largely from public transit access trips, while those in Lower East Side would draw from bike trips and those in Upper East Side would draw from taxi trips.

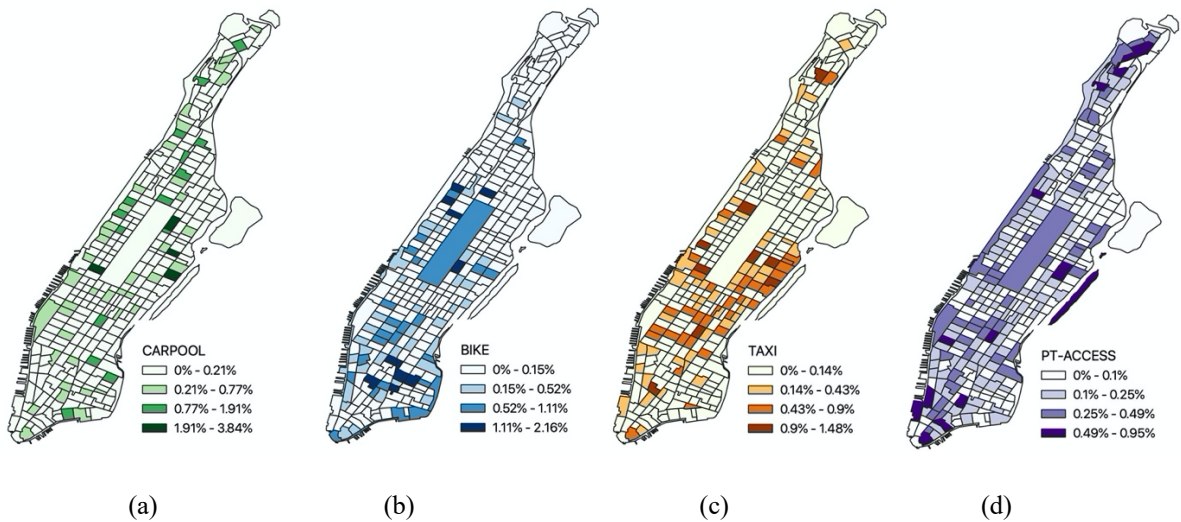

**FIGURE 8.** Substitution of (a) carpool (b) bike (c) taxi (d) public transit access trips (%) per TAZ.

## 5.4 Revenue analysis

The fare model for many e-scooter companies is consistent around $fare = \$1 + \$0.15 \times ridership\ duration(min)$ (per Lime). From the operators' perspective such as Lime and Bird, we can determine a potential average revenue by providing the e-scooter service in Manhattan. With the 75K total trips in Manhattan—supposing an average trip distance of 1.6 miles reflecting about 12 minutes (BCDOT, 2018)—approximately $210K daily revenue or $77M annual revenue is predicted with a supply of 2K e-scooters.

Figure 9 shows the relationship between the distance traveled and the total revenue from each travel mode. As the distance traveled increases, the total revenue decreases, as a result from direct substitution by e-scooters. One interesting fact is that many residents in Manhattan prefer walking short distances. Therefore, although the substitution rate of walking trips is not as high as the other modes, the revenue from replacements of short distance walking trips takes a large portion of the total.

The other factor revealed from the model estimation is access/egress trips to public transit within Manhattan. The average value of $F'_{pt,i}$ is 0.069. The small number makes sense since public transit trips are much larger in Manhattan per zone. The zonal distribution of e-scooter trips substituting access/egress to public transit can be found in Figure 8(d).

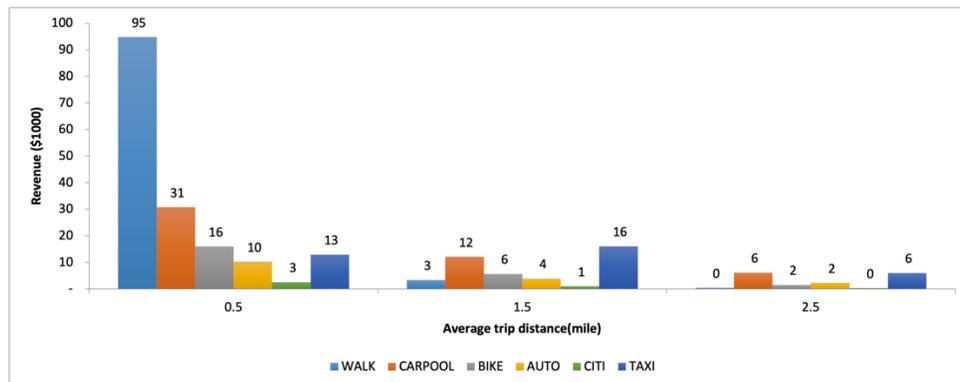

(a)

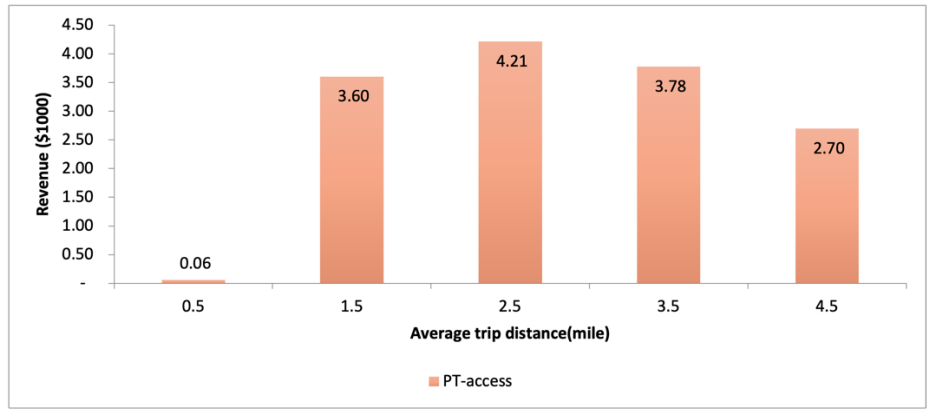

(b)

**FIGURE 9.** The revenue from substituted modes(a) and access trips to public transit(b) by distance.

The revenue structure for access trips differs from the taxi substitution trips because the former's travel time generally remains around the same regardless of the length of the public transit trip. Reported from the 2011 RHTS data, the average access time to public transit is 5.13 minutes (0.085 hr). We see in Figure 9 that there is an insignificant change in the revenues from substitution of access/egress trips when compared to the other trips in relation to distance. A future consideration is to include public transit trips entering and leaving Manhattan; we expect to see a similar revenue pattern even with longer distances.

## 5. CONCLUSIONS

We conduct the first demand forecasting model study for e-scooter mobility to explain which modes the trips could substitute. The demand model is based on data available from a pilot conducted in Portland, Austin and Chicago. The model is applied to Manhattan data to find a potential market of 75K daily trips and $77M annual revenue assuming 12-minute trips.

In addition to the demand model, we propose and estimate an exploratory multifactor model to explain where the predicted trips may come from. At 5% significance level, the e-scooter trips are shown to substitute trips from biking, carpool, and taxi, and access trips to public transit. In addition, the carpool trips are three times more likely to be substituted than the bike trips and taxi trips. E-scooters are likely to compete with the other modes at shorter distances than at longer distances, with the substitution rate($P_d$) dropping by an order of magnitude from 0.5 mile distances to 5.5 miles.

The estimated parameters allow us to analyze the spatial distribution for the modes and break down those substituted trips by distance. As a result, we see a clear distinction between the distance structure of substituted direct trips and that of access/egress trips for public transit. Having the fleet size variable, we can also analyze the elasticity of revenue or taxi trip substitution with respect to it.

The research results above are promising to give insights to government or city transportation planners. With the growing popularity of environment-friendly-transportation, e-scooters are certainly a sustainable mode providing significantly less pollution. The economical demand analysis we proposed can promote the development of policy and infrastructure relating to e-scooter systems. For example, e-scooters are driving much of the discussion on curb management practices (Sisson, 2018) as well as mobility data sharing (Carey, 2020) and as a core part of Mobility-as-a-Service platforms (Simlett and Møller, 2020).

For future research, there are some constraints to overcome. There are potential changes in individual activity patterns (scheduling, trip chaining) as well as in travel demand with a newly added service. Quantitative variables that could affect the likelihood of riding e-scooters such as weather, traffic or road conditions can also be considered. Furthermore, we can consider the effect of diverse trip attributes including trip purpose, traffic infrastructure like bike/e-scooter lanes, and safety policy. Applying a spatial correlation can be one of the future directions. The e-scooter demand in adjacent regions can be correlated with that of a target zone, implying how convenient e-scooter trips are in corresponding neighborhoods which may share common urban environmental factors. Nonlinear distance factors can be considered for modes like public transit. Capturing non-commuter behavior would also help. In our study, only Citi Bike data captures tourist trips. In the Portland pilot, a substantial proportion of e-scooter riders were non-Portlanders. According to its survey data, 48% of visitors' e-scooter trips had been replaced from motorized travel modes such as taxi, FHV, and rental cars (PBOT, 2018).

**ACKNOWLEDGMENTS**

This research was conducted with support from the C2SMART University Transportation Center (USDOT #69A3551747124).

**AUTHOR CONTRIBUTIONS**

The authors confirm contribution to the paper as follows: study conception and design: ML, GY, YH, JC; data collection: ML, YH; analysis: ML; interpretation of results: ML, GY, JC; draft manuscript preparation: ML, JC, GY.